\documentclass[12pt]{article}
\usepackage{amsfonts,amssymb,amsmath}
\usepackage[dvips]{epsfig}
\newcounter{tempeq}
\begin{document}
\title{Approach to a Parity Deformed Jaynes-Cummings Model and the Maximally Entangled States}
\author{A. Dehghani$^{1}$\thanks{Email: a\_dehghani@tabrizu.ac.ir,  alireza.dehghani@gmail.com},
 \hspace{1mm}B. Mojaveri$^{2}$\thanks{Email: bmojaveri@azaruniv.ac.ir; bmojaveri@gmail.com}\hspace{2mm}
  ,\hspace{1mm} S. Shirin$^{1}$\thanks{Email: siminshirin2000@gmail.com}\hspace{1mm} and\hspace{1mm} S. Amiri$^{2}$\thanks{Email: s.amiri@azaruniv.edu} \\
{\small {\em $^{1}$Department of Physics, Payame Noor University,
P.O.Box 19395-3697 Tehran, I.R. of Iran\,}}\\
{\small {\em $^{2}$Department of Physics, Azarbaijan Shahid Madani
University, PO Box 51745-406, Tabriz, Iran \,}}} \maketitle
\begin{abstract}
A parity deformed Jaynes-Cummings model (JCM) is introduced, which describes an interaction of a two-level atom with a $\lambda$-deformed quantized field. In the rotating wave approximation (RWA), all eigen-values and eigen-functions of this model are obtained exactly. Assuming that initially the field is prepared in the Wigner cat state (WCS) and the two-level atom is in the excited state, it has been shown that the atomic Rabi oscillations exhibit a quasi-periodic behavior in the collapse and revival patterns. The influence of the deformation parameter on the time evolution of non-classical features of the radiation field such as the sub-Poissonian statistics and squeezing effect are also analyzed. Interestingly, the main finding here is that we can realize maximally entangled atom-field states. In this note it is shown that the high fidelity is possible in the weak coupling regime, while the deformation parameter becomes large values.

\end{abstract}
\section{Introduction}
The Jaynes-Cummings model (JCM) which is extensively used in quantum optics describes the interaction of a single quantized radiation field with a two-level atom \cite{Jaynes}
\begin{eqnarray}
&&\hspace{-2.2cm}H_{JC}=\omega \left(a^{\dag} a+\frac{1}{2}\right)\sigma_{0} +\frac{\omega_{0}}{2}\sigma_{3}+g(a^{\dag}\sigma_{-}+a\sigma_{+})
\end{eqnarray}
where $a^{\dag}$ and $a$ are the photon creation and annihilation operators and satisfy the boson oscillator algebra, i.e. $[a, a^{\dag}]=1$. The spin operators $\sigma_{\pm}=\frac{1}{2}(\sigma_{1}\pm i\sigma_{2})$, with $\sigma_{1}$ , $\sigma_{2}$ and $\sigma_{3}$ being the Pauli matrices and $\sigma_{0}$ is the identity matrix. Here, $g$ is a coupling constant, $\omega$ is the radiative field mode frequency and $\omega_{0}$ the atomic frequency. The solvability and applications of this model has long been discussed in Refs. \cite{Jaynes, Narozhny}. This simple model describes various quantum mechanical phenomena, for example, Rabi oscillations \cite{Narozhny, aga}, collapse and revivals of the atomic population inversion \cite{Cummings} and entanglement between atom and field \cite{Phoenix}. Furthermore, JCM is one of several possibility schemes of producing the nonclassical states \cite{Short}. The dynamics predicted by the JCM has been proven in experiments with Rydberg atom in high quality cavities \cite{Raimond}. Since the JCM is an ideal model in quantum optics, it's various extensions such as intensity dependent coupling, two photons or multi-photon transitions, two or three cavity modes for three-level atoms and the Tavis-Cummings model, which describes the interaction between a quantized field and a group of two-level atoms have been proposed \cite{Singh, Tav}. In 1984 Sukumer and Buck studied the above models by using algebraic operator methods \cite{Sukumar}. On the other hand, the supergroup theoretical approach to JCM leads to the exact solvability of this model and the representation theory of super-algebras \cite{Buzano}. More recently, it was found by many authors that the ordinary creation and annihilation operators in the JCM may be replaced by the q-deformed partners, namely, the $q$-deformed JCM \cite{chaichian}. Later on, the JCM has been adopted with a Kerr nonlinearity within the framework of $f$-oscillator formalism \cite{Chang}. Furthermore, the investigations of a class of shape-invariant bound state problem, which represents a two-level system, leads to the generalized JCM \cite{Aleixo2}.

Besides the above generalizations, in the recent years a lot of interest has been performed to extension and deformation of the boson oscillator algebra. One of the most interesting algebra which is not related to the $q$(or $f$) -calculus is $\lambda-$ deformed (or Wigner) algebra as an obvious modification of the Heisenberg algebra \cite{Wigner, Yang}. According to Wigner's new quantization method, the Wigner-Heisenberg algebra (WHA) is raised as a unital algebra with the generators $\{1, \mathfrak{a}, \mathfrak{a}^{\dag}, \hat{R}\}$, which satisfy the (anti-)commutation relations
\begin{eqnarray}
&&\hspace{-14mm}[\mathfrak{a}, \mathfrak{a}^{\dag}]=1+2\lambda \hat{R},\hspace{5mm}\{\hat{R}, \mathfrak{a}\}=\{\hat{R},\mathfrak{a}^{\dag}\}=0.\end{eqnarray}
Here $\lambda$ is a real positive constant called Wigner parameter and $\hat{R}$ is Hermitian and unitary operator also possessing the following properties\\
\begin{eqnarray}
&&\hspace{-14mm}\hat{R}^{2}=I, \hat{R}^{\dag}=\hat{R}^{-1}=\hat{R}.\end{eqnarray}
This operator acts in the Hilbert space of eigenfunctions as:
\begin{eqnarray}
&&\hspace{-14mm}\hat{R}|n\rangle= (-1)^n|n\rangle,\end{eqnarray}
which means that $\hat{R}$ commutes with number operator $N$ that includes the eigenvector $|n\rangle$\footnote{The Fock state $|n\rangle$ recalls the generalized Hermite polynomial which reduces to the ordinary ones, while $\lambda$ tends to zero \cite{Szego}.}, such that $N|n\rangle= n|n\rangle$. The number operator $N$ is in general different from the product $\mathfrak{a}^{\dag}\mathfrak{a}$, and postulated to satisfy the following relations:
\renewcommand\theequation{\arabic{tempeq}\alph{equation}}
\setcounter{equation}{0} \addtocounter{tempeq}{5}
\begin{eqnarray}
&&\hspace{-14mm}[N, \mathfrak{a}]=-\mathfrak{a}, [N, \mathfrak{a}^{\dag}]=\mathfrak{a}^{\dag},\\
&&\hspace{-10mm}\mathfrak{a}^{\dag}\mathfrak{a}=N+\lambda(1-\hat{R}).\end{eqnarray} Which provide us with the following irreducible representation of WHA: \renewcommand\theequation{\arabic{tempeq}\alph{equation}}
\setcounter{equation}{0} \addtocounter{tempeq}{1}
\begin{eqnarray}&&\hspace{-14mm}\mathfrak{a}|2n\rangle=\sqrt{2n}|2n-1\rangle,\hspace{23mm}\mathfrak{a}|2n+1\rangle=\sqrt{2n+2\lambda+1}|2n\rangle,\\
&&\hspace{-14mm}\mathfrak{a}^{\dag}|2n\rangle=\sqrt{2n+2\lambda+1}|2n+1\rangle,\hspace{5mm}\mathfrak{a}^{\dag}|2n+1\rangle=\sqrt{2n+2}|2n+2\rangle.\end{eqnarray}
It is clear that the above representation is really different than the $f$- deformed realization of the Heisenberg algebra \cite{Vogel}. Therefore, the introduced $\lambda-$deformed algebra in (5) can be considered as a new deformation of the simple harmonic oscillator with significant features in quantum optics \cite{Sage, Dehgan, Dehgan2}.

Due to the physical significance of deformed JCM in quantum optics on the one hand, and the central role of the parity operator in the theory of deformation on the other, we then generalize the well-known JCM to a parity deformed-Hermitian case in terms of $\lambda-$deformed boson operators. These Hermitian operators arise from a special deformation of canonical bosonic commutation relations, allowing us a mathematically rigorous treatment of our deformed interaction Hamiltonian and extracting the energy spectrum and the corresponding eigen-vectors. Preparing the initial field in the $\lambda-$ deformed cat states, we will investigate on the collapse and revival phenomena in the Rabi oscillations of the atomic inversion. By setting the deformation parameter, detuning, coupling constant and average photon number of the field statistics, the fidelity and the degree of entanglement of the atom-field states may be adjusted.

The organization of the paper is as follows: We start by introducing a parity deformed JCM and its solutions in section 2. The time evolution of the system is considered in section 3. We study atomic dynamics in Section 4 and discuss some relevant physical phenomenon of collapses and revivals. Sections 5 and 6 are devoted to the calculation of the fidelity and the degree of entanglement in which the numerical results and their discussions are presented. We study nonclassical properties of the atomic system, namely, in Section 7, we evaluate the Mandel's $Q$ parameter, the normal squeezing of the field. The paper is concluded in section 8 with a brief conclusion.
\section{Parity deformation of JCM}
We begin by introducing the $\lambda-$deformed Hamiltonian, describes an interaction between a two-level atom driven by a $\lambda-$deformed quantized field, as follows
\renewcommand\theequation{\arabic{tempeq}\alph{equation}}
\setcounter{equation}{-1} \addtocounter{tempeq}{1}
\begin{eqnarray}
&&\hspace{-2.2cm}H_{\lambda}=\frac{\omega}{2}\{\mathfrak{a},{\mathfrak{a}}^{\dag}\}+\frac{\omega_{0}}{2}\sigma_{3}+g({\mathfrak{a}}^{\dag}\sigma_{-}+\mathfrak{a}\sigma_{+}),
\end{eqnarray}
which generalizes the ordinary JC Hamiltonian in the rotating-wave approximation. It reduces to the well-known JCM [Eq. (1)], while the annihilation and creation operators are those associated with the harmonic oscillator. We also note that the Hamiltonian $H_{\lambda}$ is super-symmetric when $\omega=\omega_{0}$ (exact resonance) and $g=0$ (absence of coupling). Along with substitution Eq. (2) in Eq. (7), we can recast the Hamiltonian $H_{\lambda}$ into
\renewcommand\theequation{\arabic{tempeq}\alph{equation}}
\setcounter{equation}{-1} \addtocounter{tempeq}{1}
\begin{eqnarray}
&&\hspace{-2.2cm}H_{\lambda}=\omega\left({\mathfrak{a}}^{\dag}\mathfrak{a}+\frac{1}{2}+\lambda\hat{R}\right)+\frac{\omega_{0}}{2}\sigma_{3}+g({\mathfrak{a}}^{\dag}\sigma_{-}+\mathfrak{a}\sigma_{+}).\end{eqnarray}
In this case the model possesses an exact solution because of existence of an integral of motion, ${\mathfrak{a}}^{\dag}\mathfrak{a}+\frac{1}{2}\sigma_{3}$, which commutes with the Hamiltonian $H_{\lambda}$ and allows us to decompose all the representation space of the atom-field system as the tensor product of the Hilbert space associated to the field, $\mathcal{H}_{\lambda}$, times the Hilbert space associated to the spin, $\mathcal{H}_{\emph{f}}$,
\renewcommand\theequation{\arabic{tempeq}\alph{equation}}
\setcounter{equation}{-1} \addtocounter{tempeq}{1}
\begin{eqnarray}
&&\hspace{-2.2cm}\mathcal{H}:=\mathcal{H}_{\lambda}\otimes \mathcal{H}_{\emph{f}}=\left\{|2n,+\rangle=\left(\begin{array}{c}|2n\rangle \\0 \\\end{array}\right),|2n+1,-\rangle=\left(\begin{array}{c}0 \\|2n+1\rangle \\\end{array}\right)\right\}^{\infty}_{n=0},
\end{eqnarray}
or
\renewcommand\theequation{\arabic{tempeq}\alph{equation}}
\setcounter{equation}{-1} \addtocounter{tempeq}{1}
\begin{eqnarray}
&&\hspace{-2.2cm}\left\{|2n+1,+\rangle=\left(\begin{array}{c}|2n+1\rangle \\ 0 \\\end{array}\right),|2n+2,-\rangle=\left( \begin{array}{c}0 \\|2n+2\rangle \\\end{array}\right)\right\}^{\infty}_{n=0}.\end{eqnarray}
Here, $|2n,+\rangle$ is the state in which the atom is in the excited state $|+\rangle$ and the field has $2n$ photons, and a similar description holds for the state $|2n+1,-\rangle$, where $|-\rangle$ is the atom ground state. Using the Fock space $\mathcal{H}$ given in (9), we can find the following matrix representation of the $\lambda-$deformed JC Hamiltonian $H_{\lambda}$:
\renewcommand\theequation{\arabic{tempeq}\alph{equation}}
\setcounter{equation}{-1} \addtocounter{tempeq}{1}
\begin{eqnarray}
&&\hspace{-2.2cm}H_{\lambda}=\left(\begin{array}{cc} \omega\left(2n+\lambda+\frac{1}{2}\right)+\frac{\omega_{0}}{2} & g\sqrt{2n+2\lambda+1}\\g\sqrt{2n+2\lambda+1} & \omega\left(2n+\lambda+\frac{3}{2}\right)-\frac{\omega_{0}}{2} \\ \end{array}\right).\end{eqnarray}
It is easy to see that the normalized energy eigen-states of $H_{\lambda}$ are
\renewcommand\theequation{\arabic{tempeq}\alph{equation}}
\setcounter{equation}{0} \addtocounter{tempeq}{1}
\begin{eqnarray}
&&\hspace{-2.2cm}|E^{+}_{n}\rangle=c_{1}|2n,+\rangle+c_{2}|2n+1,-\rangle,\\
&&\hspace{-2.2cm}|E^{-}_{n}\rangle=c_{2}|2n,+\rangle-c_{1}|2n+1,-\rangle,
\end{eqnarray}
where the coefficients $c_{1(2)}$ are given by:
\renewcommand\theequation{\arabic{tempeq}\alph{equation}}
\setcounter{equation}{0} \addtocounter{tempeq}{1}
\begin{eqnarray}
&&\hspace{-2.2cm}c_{1}=\frac{\Delta-\Omega_{n, \lambda}}{\sqrt{(\Delta-\Omega_{n, \lambda})^2+4g^{2}(2n+2\lambda+1)}},\\
&&\hspace{-2.2cm}c_{2}=\frac{2g\sqrt{2n+2\lambda+1}}{\sqrt{(\Delta-\Omega_{n, \lambda})^2+4g^{2}(2n+2\lambda+1)}},
\end{eqnarray}
in which $\Delta(=\omega-\omega_{0})$ and $\Omega_{n, \lambda}(=\sqrt{\Delta^2+4g^{2}(2n+2\lambda+1)})$ are defined as detuning parameter and a generalized Rabi frequency, respectively. The energy eigenvalues corresponding to the eigen-states in Eqs. (12) are
\renewcommand\theequation{\arabic{tempeq}\alph{equation}}
\setcounter{equation}{-1} \addtocounter{tempeq}{1}
\begin{eqnarray}
&&\hspace{-2.2cm}E^{\pm}_{n, \lambda}=(2n+\lambda+1)\omega\pm\frac{\Omega_{\lambda}}{2}.
\end{eqnarray}
\begin{figure}
\begin{center}
\epsfig{figure=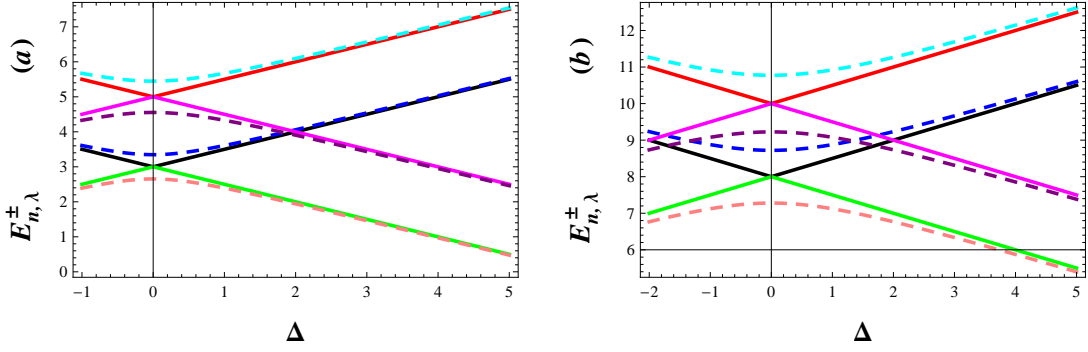,width=15cm}
\end{center}
\caption{\footnotesize Dependence of eigenvalues $E^{\pm}_{n, \lambda}$ on detuning $\Delta$. The continuous curve corresponds to $g = 0. 01$. The dashed curves (a) and (b) corresponds to $(g = 0$, $\lambda = 0)$ and $(g = 0$, $\lambda = 50)$, respectively. The dashed curves with positive and negative slopes correspond respectively to $E^{+}_{n, \lambda}$ and $E^{-}_{n, \lambda}$ . Lower part of the figure is for $n = 1$ and upper part for $n = 2$.}
\end{figure}
The energy difference between the levels $E^{+}_{n, \lambda}$ and $E^{-}_{n, \lambda}$ is $\Omega_{n, \lambda}$. The minimum of the separation occurs when $\Delta$ equals to zero and the corresponding difference is $2g\sqrt{2n+2\lambda+1}$. In Fig. 1 (a) and (b), respectively, we have plotted the energy eigenvalues $E^{+}_{n, \lambda}$ and $E^{-}_{n, \lambda}$ as functions of $\Delta$ for given values of $\lambda= 0, 50$. The dotted lines represent the eigenvalues when $g=0$, i. e. $E^{\pm}_{n, \lambda}=(2n+\lambda+1)\omega\pm\frac{\Delta}{2}$. In this case the eigenvalues cross each other as increases from negative to positive values. The continuous lines represent the energy eigenvalues for $g= 0.01$. The diverging eigenvalue separation beyond the minimum separation indicates level repulsion in the eigenvalues of the dressed atom. As Figs. 1(a) and 1(b) show, the repulsion between energy levels increases while the deformation parameter $\lambda$ gets bigger. Where the latter, also, leads to shift the energy levels to the positive side.
\section{Evolution Of Atom-Filed State}
In order to study the influence of the deformation on the dynamics of the system, firstly we decompose the Hamiltonian (8) as follows
\renewcommand\theequation{\arabic{tempeq}\alph{equation}}
\setcounter{equation}{-1} \addtocounter{tempeq}{1}
\begin{eqnarray}
&&\hspace{-2.2cm}H_{\lambda}=H_{0}+H^{'},
\end{eqnarray}
where
\renewcommand\theequation{\arabic{tempeq}\alph{equation}}
\setcounter{equation}{0} \addtocounter{tempeq}{1}
\begin{eqnarray}
&&\hspace{-2.2cm}H_{0}=\omega\left({\mathfrak{a}}^{\dag}\mathfrak{a}+\frac{1}{2}+\lambda\hat{R}\right)+\frac{1}{2}\omega_{0}\sigma_{3},\\
&&\hspace{-2.2cm}H^{'}=g({\mathfrak{a}}^{\dag}\sigma_{-}+\mathfrak{a}\sigma_{+}).\end{eqnarray}
In the interaction picture generated by $H_{0}$, the Hamiltonian of the system can be written as
\renewcommand\theequation{\arabic{tempeq}\alph{equation}}
\setcounter{equation}{-1} \addtocounter{tempeq}{1}
\begin{eqnarray}&&\hspace{-2.2cm}H_{I}=e^{iH_{0}t}H'e^{iH_{0}t}=g\left({\mathfrak{a}}^{\dag}\sigma_{-}e^{i\Delta t}+\mathfrak{a}\sigma_{+}e^{-i\Delta t}\right).\end{eqnarray}
We now proceed to solve the equation of motion of this system in an interaction picture, i.e.
\renewcommand\theequation{\arabic{tempeq}\alph{equation}}
\setcounter{equation}{-1} \addtocounter{tempeq}{1}
\begin{eqnarray}
&&\hspace{-2.2cm}H_{I}\Psi(t)=i\frac{\partial}{\partial t}\Psi(t).\end{eqnarray}
At any time $t$, the wave function $\Psi(t)$ is expanded in terms of the states $|2n,+\rangle$ and $|2n+1,-\rangle$ as follows
\renewcommand\theequation{\arabic{tempeq}\alph{equation}}
\setcounter{equation}{-1} \addtocounter{tempeq}{1}
\begin{eqnarray}
&&\hspace{-2.2cm}\Psi(t)=\sum^{\infty}_{n=0}[c_{+,2n}(t)|2n,+\rangle+c_{-,2n+1}(t)|2n+1,-\rangle]
\end{eqnarray}
Clearly, $\Psi(t)$ is determined completely once the coefficients $c_{+,2n}(t)$ and $c_{-,2n+1}(t)$ are known. Inserting (19) into (18) we obtain the following general solution for the probability amplitudes, $c_{+,2n}(t)$ and $c_{-,2n+1}(t)$, as:
\renewcommand\theequation{\arabic{tempeq}\alph{equation}}
\setcounter{equation}{0} \addtocounter{tempeq}{1}
\begin{eqnarray}
&&\hspace{-2.2cm}c_{+,2n}(t)=\left\{c_{+,2n}(0)\left[\cos\left(\frac{\Omega_{n,\lambda}}{2}t\right)+i\frac{\Delta}{\Omega_{n,\lambda}}\sin\left(\frac{\Omega_{n,\lambda}}{2}t\right)\right]\right.\nonumber\\
&&\hspace{2.2cm}\left.-2ig\frac{\sqrt{2n+2\lambda+1}}{\Omega_{n,\lambda}}c_{-,2n+1}(0)\sin\left(\frac{\Omega_{n,\lambda}}{2}t\right)\right\}e^{-i\frac{\Delta}{2} t},\\
&&\hspace{-2.2cm}c_{-,2n+1}(t)=\left\{c_{-,2n+1}(0)\left[\cos\left(\frac{\Omega_{n,\lambda}}{2}t\right)-i\frac{\Delta}{\Omega_{n,\lambda}}\sin\left(\frac{\Omega_{n,\lambda}}{2}t\right)\right]\right.\nonumber\\
&&\hspace{2.2cm}\left.-2ig\frac{\sqrt{2n+2\lambda+1}}{\Omega_{n,\lambda}}c_{+,2n}(0)\sin\left(\frac{\Omega_{n,\lambda}}{2}t\right)\right\}e^{i\frac{\Delta}{2} t},\end{eqnarray}
where $c_{-,2n+1}(0)$ and $c_{+,2n}(0)$ are determined from the initial conditions of the system, which is supposed initially in its excited state, i.e. $c_{+,2n}(0)=c_{2n}(0)$ and $c_{-,2n+1}(0) = 0$. Here the initial condition for the field is described by $c_{2n}(0)$. For this case in particular, we have
\renewcommand\theequation{\arabic{tempeq}\alph{equation}}
\setcounter{equation}{0} \addtocounter{tempeq}{1}
\begin{eqnarray}
&&\hspace{-2.2cm}c_{+,2n}(t)=c_{2n}(0)\left[\cos\left(\frac{\Omega_{n,\lambda}}{2}t\right)+i\frac{\Delta}{\Omega_{n,\lambda}}\sin\left(\frac{\Omega_{n,\lambda}}{2}t\right)\right]e^{-i\frac{\Delta}{2} t},\\
&&\hspace{-2.2cm}c_{-,2n+1}(t)=-2ig\frac{\sqrt{2n+2\lambda+1}}{\Omega_{n,\lambda}}c_{2n}(0)\sin\left(\frac{\Omega_{n,\lambda}}{2}t\right)e^{i\frac{\Delta}{2} t},\end{eqnarray}
This set of equations gives us the solution of the problem. In order to calculate some physical quantities of interest, we need only to specify the initial photon number distribution of the field $|c_{2n}(0)|^{2}$.

The field we are considering in this work is being treated as an $\lambda$-deformed oscillator, we have several options to describe it. We focus on the situations which the field as an eigen-state of the $\lambda$-deformed annihilation operator, introduced in Ref. \cite{Dehgan}, i.e. $\mathfrak{a}^2|W\rangle_{\lambda, +}=w^2|W\rangle_{\lambda, +}$, and it's number state expansion is\footnote{Here, $I_{\lambda}(x)$ refers to the modified Bessel function of the first type \cite{abramo}, with a convergency radius of infinity that has been used in order to normalize the WCSs to unity i.e. $_{\lambda, +}\langle W|W\rangle_{\lambda, +}=1$ for $w\in \mathbb{C}$. It is worth mentioning that the states $|W\rangle_{\lambda, +}$ is defined for $\lambda>-\frac{1}{2}$.}
\renewcommand\theequation{\arabic{tempeq}\alph{equation}}
\setcounter{equation}{-1} \addtocounter{tempeq}{1}
\begin{eqnarray}&&\hspace{-1.5cm}|w\rangle_{\lambda, +}:=\sqrt{\frac{\left(\frac{|w|}{\sqrt{2}}\right)^{2\lambda-1}}{I_{\lambda-\frac{1}{2}}(|w|^2)}}\sum^{\infty}_{n=0}{\frac{w^{2n}}{\sqrt{2^{2n}n!\Gamma(n+\lambda+\frac{1}{2})}}}|2n\rangle.\end{eqnarray}Therefore, in this case, we have
\renewcommand\theequation{\arabic{tempeq}\alph{equation}}
\setcounter{equation}{-1} \addtocounter{tempeq}{1}
\begin{eqnarray}&&\hspace{-1.5cm}|c_{2n}(0)|^{2}={\frac{{\left(\frac{|w|}{\sqrt{2}}\right)}^{4n+2\lambda-1}}{{n!\Gamma(n+\lambda+\frac{1}{2})I_{\lambda-\frac{1}{2}}(|w|^2)}}}.
\end{eqnarray}
It is worth to mention that, as a second option, one may choose the initial state of the field as the Wigner negative binomial states already studied in Ref. \cite{Dehgan2}. They are mathematically equivalent to those nonlinear ones and may be expected to bring new quantum features.
\section{Atomic Dynamics}
In this section, we are interested in studying the temporal evolution of atomic inversion. Which, in turn, is specified by the expectation value of the inversion operator as:
\renewcommand\theequation{\arabic{tempeq}\alph{equation}}
\setcounter{equation}{-1} \addtocounter{tempeq}{1}
\begin{eqnarray}&&\hspace{-1.5cm}\langle \sigma_{z}\rangle=\sum^{\infty}_{n=0}\left(|c_{+,2n}(t)|^2-|c_{-,2n+1}(t)|^2\right).
\end{eqnarray}
Substituting Eqs. (21) and (23) into (24), we obtain, for an atom prepared initially in the excited state,
\renewcommand\theequation{\arabic{tempeq}\alph{equation}}
\setcounter{equation}{-1} \addtocounter{tempeq}{1}
\begin{eqnarray}&&\hspace{-1.5cm}\langle \sigma_{z}\rangle=\sum^{\infty}_{n=0}|c_{2n}(0)|^{2}\left[\left(\frac{\Delta}{\Omega_{\lambda}}\right)^2+(8n+8\lambda+4)\left(\frac{g}{\Omega_{\lambda}}\right)^2\cos\left(\Omega_{\lambda}t\right)\right].
\end{eqnarray}
\begin{figure}
\begin{center}
\epsfig{figure=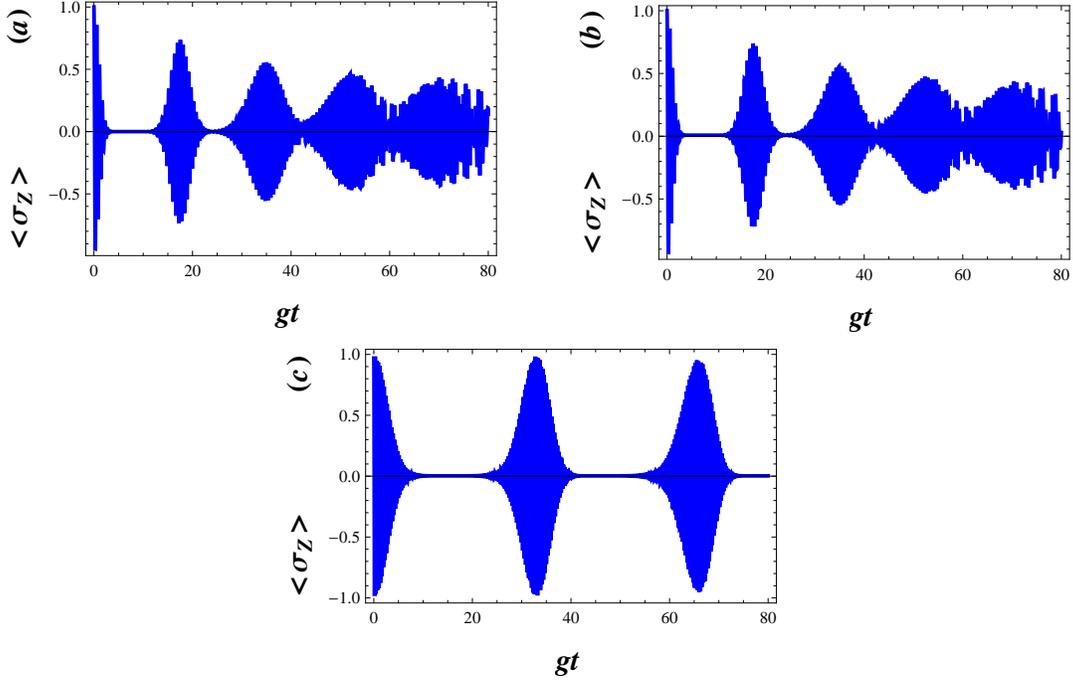,width=15cm}
\end{center}
\caption{\footnotesize Temporal evolution of the atomic inversion $\langle \sigma_{z}\rangle$ for the field initially prepared in WCS, $|w\rangle_{\lambda, +}$, with $|w|^2=30$ and $g= 0.01$. The parameters are (a):$\lambda=0$, $\Delta=0$, (b):$\lambda=0$, $\Delta=0.01$ and (c):$\lambda=50$, $\Delta=0.01$.}
\end{figure}
\begin{figure}
\begin{center}
\epsfig{figure=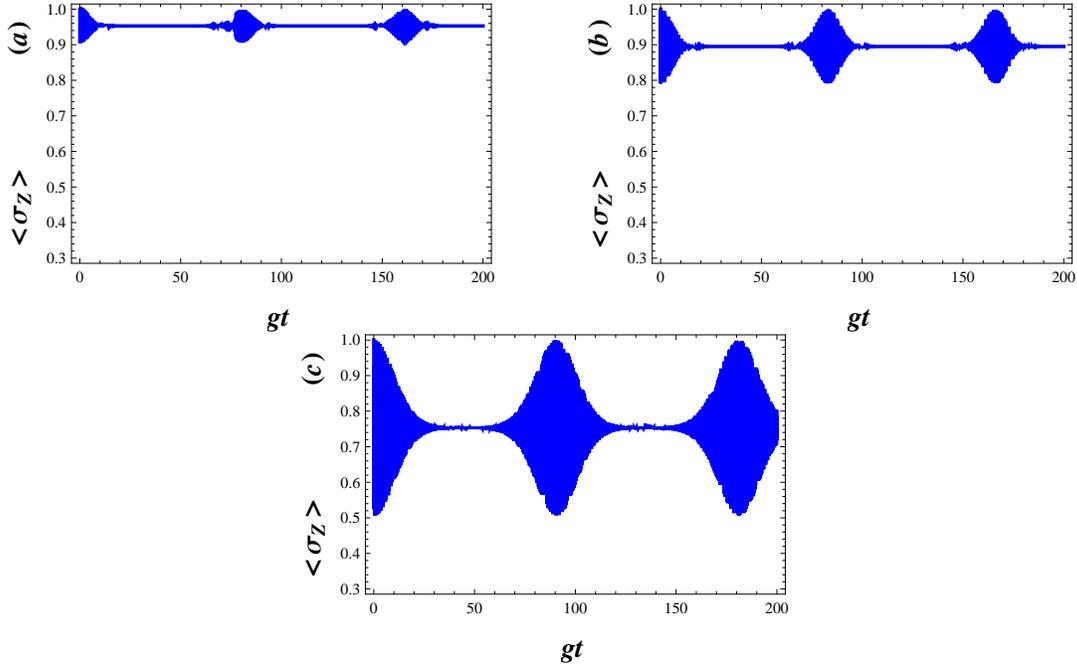,width=15cm}
\end{center}
\caption{\footnotesize Temporal evolution of the atomic inversion with $|w|^2= 30$, $g= 0.01$ and $\Delta= 0.5$. The parameters equal to (a):$\lambda= 0$, (b): $\lambda=30$ and (c): $\lambda=100$.}
\end{figure}
Numerical results of the atomic inversion when the field is in a standard cat state (i.e. when $\lambda=0$), with $(|w|^2=)30$ photons on average and detuning factors 0 and 0.01, were shown versus the scaled time $gt$ in figures 2 (a) and (b), respectively. The temporal evolution of the atomic inversion $\langle \sigma_{z}\rangle$ reveals significant discrepancies of the well-known phenomenon of collapses and revivals \cite{Bayfield}. Recall that the collapse, i.e. when the envelope of the oscillations collapses to zero, is due to the destructive interference among the probability amplitudes at different Rabi frequencies, $\Omega_{n,\lambda}$, for different photon number eigen-states. At the revival times, on the other hand, constructive interference occurs. This phenomenon also takes place when the initial field state is a WCS. In figure 2 (c) the function $\langle \sigma_{z}\rangle$ is plotted for the value $\lambda=50$. In this case, $\langle \sigma_{z}\rangle$ exhibits quasi periodic behaviour very similar to the atomic inversion of a two-photon JCM \cite{Zubairy,Puri2,Puri3}, with an effective Hamiltonian defined as $H_{eff}=\omega\left({a}^{\dag}a+\frac{1}{2}\right)+\frac{\omega_{0}}{2}\sigma_{3}+g({{a}^{\dag}}^2\sigma_{-}+{a}^2\sigma_{+})$. However, in this case note how the structure of the oscillations is much more complex than the standard Rabi oscillations. As the detuning factor $\Delta$ increases, these structures are disappeared( see figures 3), i.e. the inhibition of the radiation decay is more transparent. It is clear that the inhibited decay even occurs in the case $\lambda=0$. This behaviour is due to the influence of the parity deformation via the generalized Rabi frequency $\Omega_{n,\lambda}$. Figures 3 (a)- (c) indicate that, with increasing $\lambda$ the inhibition decay of the excited state will be balanced.
\section{Fidelity}
We now calculate the fidelity
\renewcommand\theequation{\arabic{tempeq}\alph{equation}}
\setcounter{equation}{0} \addtocounter{tempeq}{1}
\begin{eqnarray}
&&\hspace{-3.7cm}F=|\langle \Psi(0)|\Psi(t)\rangle|^2,\end{eqnarray}
which measures the ``closeness''of the two quantum states $|\Psi(t)\rangle$ and $|\Psi(0)\rangle=|w\rangle_{\lambda, +}\otimes|+\rangle$, which indicate that $F$ is unity when these two quantum states are identical. We plot the fidelity in Figs. 4, when the deformation parameter increases the fidelity decreases but remains close to its initial value( see figures 4(a)- (d)). To obtain a fidelity around $1$, one needs to enhance the deformation parameter $\lambda$ to 100 and $gt=95$. In this case, $|\Psi(t)\rangle$ becomes minimum uncertainty state which minimize the uncertainty relation, Eq. (29) in Ref. \cite{Dehgan}.
\begin{figure}
\begin{center}
\epsfig{figure=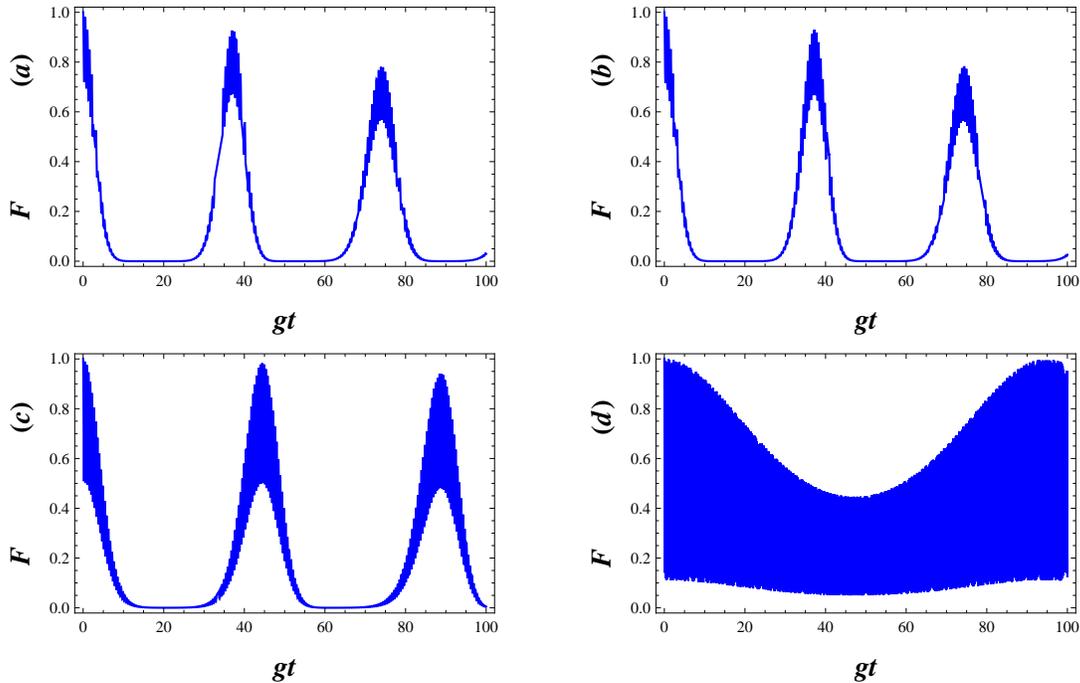,width=15cm}
\end{center}
\caption{\footnotesize Fidelity as a function of the scaled time of $gt$ in the small coupling regime with $g=0.01$ for different $\lambda(= -.25, 0, 10,100)$, other parameters are $|w|^2=9$ and $\Delta=0.1$.}
\end{figure}
\section{The von Neumann entropy}
The entropy of a radiation field is one of the fundamental problems of statistical physics. It is a very useful operational measure of the purity of the quantum state. The time evolution of the field entropy reflects the time evolution of the degree of entanglement between the atom and the field \cite{Phoenix, Gea}. We finish this section with a discussion of the Schmidt decomposition and the related von Neumann entropy as they pertain to the JCM. As this system is bipartite, a Schmidt decomposition is assured. We have already presented the solution of the time-dependent Schrodinger equation in Eq. (19), which we rewrite here as
\renewcommand\theequation{\arabic{tempeq}\alph{equation}}
\setcounter{equation}{0} \addtocounter{tempeq}{1}
\begin{eqnarray}
&&\hspace{-3.7cm}\Psi(t)=\sum^{\infty}_{n=0}[c_{+,2n}(t)|2n,+\rangle+c_{-,2n+1}(t)|2n+1,-\rangle]\nonumber,\end{eqnarray}
according to the Schmidt decomposition, for any instant in time $t$, we can always find the reduced density operator of the atom in the bare basis specified by $|+\rangle$ and $|-\rangle$ and obtain
\renewcommand\theequation{\arabic{tempeq}\alph{equation}}
\setcounter{equation}{0} \addtocounter{tempeq}{1}
\begin{eqnarray}
&&\hspace{-3.7cm}\rho_{A}=\left(\begin{array}{cc} \sum^{\infty}_{n=0}{|c_{+,2n}(t)|^2} & 0 \\0 & \sum^{\infty}_{n=0}{|c_{-,2n+1}(t)|^2} \\\end{array}\right).\end{eqnarray}
Clearly, eigenvalues of the density operator of the atom, $g_{\pm}$, can be expressed in terms of the coefficients $c_{+,2n}(t)$ and $c_{-,2n+1}(t)$ i.e. $g_{+}=\sum^{\infty}_{n=0}{|c_{+,2n}(t)|^2}$ and $g_{-}=\sum^{\infty}_{n=0}{|c_{-,2n+1}(t)|^2}$. It is easy to obtain an expression for von Neumann entropy, which for each of the subsystems of the Jaynes–Cummings model is
\renewcommand\theequation{\arabic{tempeq}\alph{equation}}
\setcounter{equation}{0} \addtocounter{tempeq}{1}
\begin{eqnarray}
&&\hspace{-3.7cm}S(\rho_{A})=-g_{+}\ln{g_{+}}-g_{-}\ln{g_{-}}.\end{eqnarray}
In figures 5, for an atom initially in the excited state and for the field initially in WCS with $|w|=3$, $g=0.01$ and $\Delta=.1$, we plot the von Neumann entropy $S(\rho_{A})$, all against the scaled time $gt$. Sometimes, the von Neumann entropy is maximum where the atom and field are nearly in maximally entangled status at that time( see Fig. 5 (c)). As the deformation parameter $\lambda$ increases, the system of the atom- field become maximally entangled (figures 5 (a)- (d)).
\begin{figure}
\begin{center}
\epsfig{figure=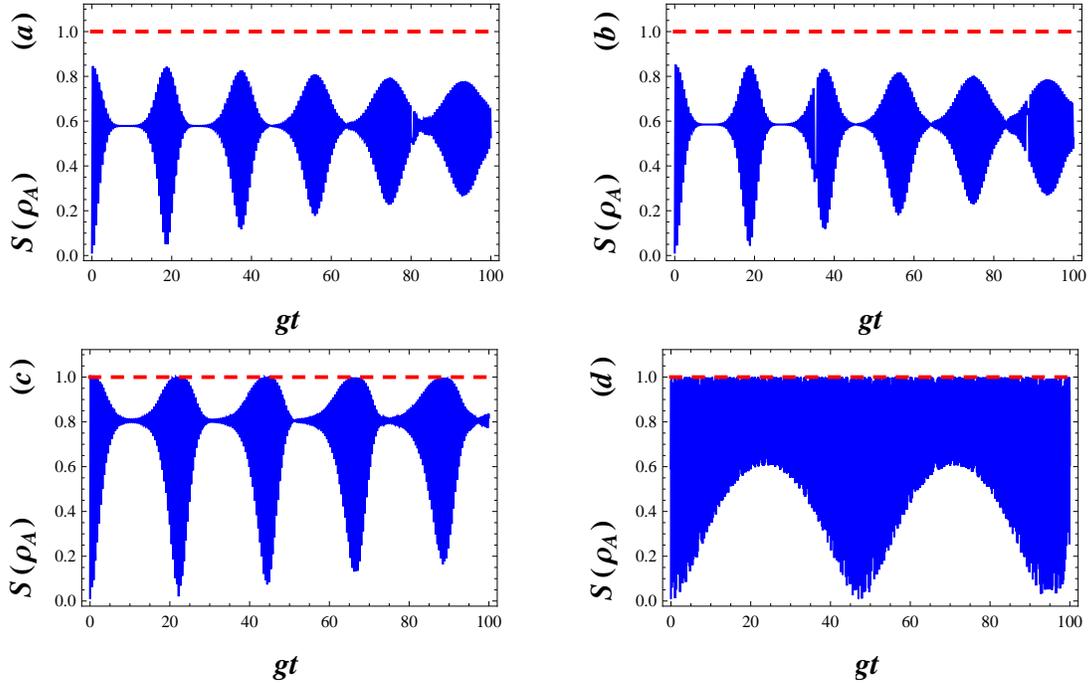,width=15cm}
\end{center}
\caption{\footnotesize Plots of entropy $S(\rho_{A})$ versus $gt$ with $g=.01, |w|^2=9,\Delta=0.1, $for various deformation parameter respectively (a):$\lambda=-0.25$, (b):$\lambda=0$, (c):$\lambda=10$ and (d):$\lambda=50$.}
\end{figure}
The essence of this subsection is summarized in Fig. 6 where we compare the von Neumann entropy and fidelity of the quantum state, $\Psi(t)$, associated with the $\lambda-$deformed JCM investigated here. Notice that entanglement between the atom and field is disappeared(i.e. the quantum state $\Psi(t)$ becomes separable), where fidelity of the quantum state passes the greatest value.
\begin{figure}
\begin{center}
\epsfig{figure=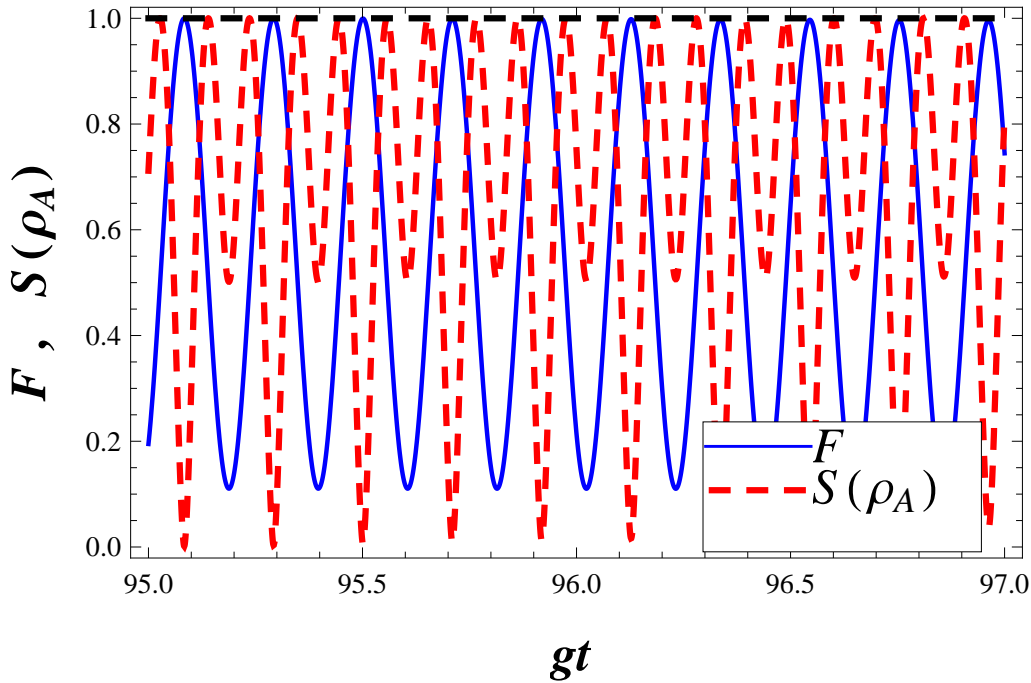,width=15cm}
\end{center}
\caption{\footnotesize The von Neumann entropy (dash) and the fidelity (solid) are plotted together. In each case a deformation parameter with $\lambda= 100, g=0.01, |w|^2=9,\Delta=0.1$ is used.}
\end{figure}
\section{Non-classical Properties}
We now examine the time evolution of the nonclassical properties of the constructed states $\Psi(t)$. To achieve this purpose, we investigate on the sub-Poissonian statistics and quadrature squeezing of them. It should be mentioned that squeezing or sub-Poissonian statistics are sufficient requirements for a state to belong the non-classical ones.
\subsection{Sub-Poisonian Statistics}
The anti-bunching effect as well as the sub-Poissonian statistics of the states $\Psi(t)$, is investigated by evaluating Mandel's $Q^{\lambda}$ parameters, which can defined as
\renewcommand\theequation{\arabic{tempeq}\alph{equation}}
\setcounter{equation}{-1} \addtocounter{tempeq}{1}
\begin{eqnarray}&&\hspace{-14mm}Q^{\lambda}=\frac{\langle{{(\mathfrak{a}^{\dag}\mathfrak{a})^2}}\rangle_{\lambda}-{\langle{{\mathfrak{a}^{\dag}\mathfrak{a}}\rangle}}_{\lambda}^{2}
}{{\langle{{\mathfrak{a}^{\dag}\mathfrak{a}}\rangle}}_{\lambda}}-1.\end{eqnarray}
\begin{figure}
\begin{center}
\epsfig{figure=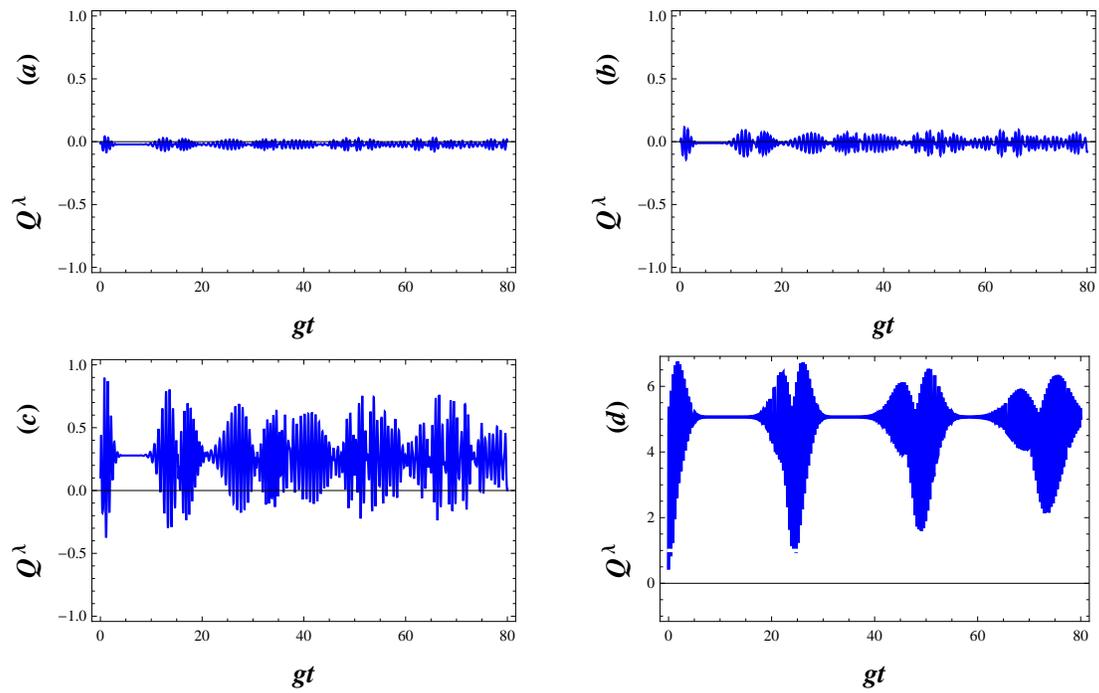,width=15cm}
\end{center}
\caption{\footnotesize Plots of the normalized variance,$Q^{\lambda}$ , as a function of the normalized time $gt$, for the field initially prepared in an annihilation operator coherent state $|w\rangle_{\lambda, +}$ with $|w|^2= 20$, $g= 0.01$ and $\Delta= 0.01$. The parameters are (a): $\lambda$= -0.25, (b):$\lambda$= 0, (c): $\lambda$= 2 and (d): $\lambda$= 10.}
\end{figure}
The inequality $Q < 0$, indicates the sub-Poissonian photon number distribution, which implies that photons are antibunched; that is, the detection of a photon makes a subsequent detection event less likely. It is well know that sub-Poissonian statistics is a signature of the quantum nature of the field. Conversely, a field for which $Q> 0$ holds is called bunched, indicating a bunching of photons. In this case, the above inequality indicates the super-Poissonian photon number distribution. Also $Q= 0$ corresponds to the canonical coherent state. Here, the angular brackets denote averaging any field operator $\hat{O}$ over an arbitrary normalizable state $\Psi(t)$ for which the mean values are well defined, i.e.
\renewcommand\theequation{\arabic{tempeq}\alph{equation}}
\setcounter{equation}{-1} \addtocounter{tempeq}{1}
\begin{eqnarray}&&\hspace{-2.2cm}\langle\hat{O}\rangle=\sum^{\infty}_{n,m}\left[c^{\ast}_{+,2m}(t)c_{+,2n}(t)\langle 2m,+|\hat{O}|2n,+\rangle\right.\nonumber\\
&&\hspace{2.2cm}\left.+c^{\ast}_{-,2m+1}(t)c_{-,2n+1}(t)\langle 2m+1,-|\hat{O}|2n+1,-\rangle\right],\end{eqnarray}
where the probability amplitudes $c_{+,2n}$ and $c_{-,2n+1}(t)$ are given by equations (21a) and (21b), respectively. In figures 7, we show the temporal evolution of the Mandel's $Q^{\lambda}$ parameters given by equation (27), when the field is initially in a $\lambda$-deformed annihilation operator coherent state introduced in (23), $|w\rangle_{\lambda, +}$, with $|w|^2=20$. One can shows that its statistics tends to fluctuate around Poissonian conduct for small values of $\lambda$( see Figs. 7(a) and 7(c)). On the other hand, the statistics of the deformed annihilation operator coherent state exhibits, in general, a super-Poissonian behaviour for the same $|w|^2$ initial value. For a larger deformation parameter, $\lambda=10$, the results are displayed in figure 7(d). In this case, the nature of oscillations is evidently different from that in figure 7 (b). In other word, its oscillatory behaviour corresponds to just a super-Poissonian statistics (figure 7 (d)). However, when the initial field state is a deformed annihilation operator state (see figures 7 (a), (c) and (d)), it is clear that the temporal evolution of the variance oscillates between sub-Poissonian and super-Poissonian statistics.
\subsection{Squeezing Effect}
Squeezing of radiation is a purely nonclassical phenomenon without any classical analogue and has attracted considerable attention owing to its low-noise property. It has been either experimentally observed or theoretically predicted in a variety of nonlinear optical processes. Now, let us consider the squeezing properties of the field by introducing the following two Hermitian field amplitudes, $\hat{x}\left(=\frac{\mathfrak{a}+{\mathfrak{a}^{\dag}}}{\sqrt{2}}\right)$ and $\hat{p}_{\lambda}\left(=\frac{\mathfrak{a}-{\mathfrak{a}^{\dag}}}{i\sqrt{2}}\right)$. The uncertainty relation for the variances of these operators are obtained as
\renewcommand\theequation{\arabic{tempeq}\alph{equation}}
\setcounter{equation}{-1} \addtocounter{tempeq}{1}
\begin{eqnarray}&&\hspace{-14mm}\langle\sigma_{xx}\rangle \langle\sigma_{pp}\rangle\geq \frac{|\langle1+2\lambda \hat{R}\rangle|^{2}}{4},\end{eqnarray}
where $\langle\sigma_{\hat{x}\hat{y}}\rangle= \frac{\langle \hat{x}\hat{y}+{\hat{y}}\hat{x}\rangle}{2}-\langle \hat{x}\rangle\langle \hat{y}\rangle$ and the angular brackets denote averaging over an arbitrary normalizable state for which the mean values are well defined, $\langle \hat{y}\rangle={\langle \Psi(t)|}\hat{y}|\Psi(t)\rangle$. It can be said that a state is squeezed if the condition $\langle\sigma_{xx}\rangle < \frac{|\langle1+2\lambda \hat{R}\rangle|}{2}$ or $\langle\sigma_{pp}\rangle < \frac{|\langle1+2\lambda \hat{R}\rangle|}{2}$ is fulfilled \cite{Walls, WODKIEWICZ}. In other words, a quantum state is called squeezed state if it has less uncertainty in one parameter ($\hat{x}$ or $\hat{p}$) than coherent state. Then to measure the degree of squeezing we introduce the squeezing factors $S_{x(p)}$ \cite{Buzek1}, corresponding with the state $\Psi(t)$, respectively
\renewcommand\theequation{\arabic{tempeq}\alph{equation}}
\setcounter{equation}{0} \addtocounter{tempeq}{1}
\begin{eqnarray}&&\hspace{-14mm}S_{x}=\frac{\langle\sigma_{xx}\rangle-\frac{|\langle1+2\lambda\hat{R}\rangle|}{2}}{\frac{|\langle1+2\lambda\hat{R}\rangle|}{2}},\\
&&\hspace{-14mm}S_{p}=\frac{\langle\sigma_{pp}\rangle-\frac{|\langle1+2\lambda\hat{R}\rangle|}{2}}{\frac{|\langle1+2\lambda\hat{R}\rangle|}{2}},\end{eqnarray}
which results that the squeezing condition takes the simple form $S^{\lambda}_{x(p),i}< 0$. By using the mean values of the generators of the WHA,
\renewcommand\theequation{\arabic{tempeq}\alph{equation}}
\setcounter{equation}{0} \addtocounter{tempeq}{1}
\begin{eqnarray}
&&\hspace{-3.7cm}\langle \mathfrak{a}\rangle=\langle{\mathfrak{a}}^{\dag}\rangle=0\\
&&\hspace{-3.7cm}\langle{\mathfrak{a}}^{2}\rangle=\sum^{\infty}_{n}\left[c^{\ast}_{+,2n}(t)c_{+,2n+2}(t)\sqrt{(2n+2)(2n+2\lambda+1)}\right.\nonumber\\
&&\hspace{7mm}\left.+c^{\ast}_{-,2n+1}(t)c_{-,2n+3}(t)\sqrt{(2n+2)(2n+2\lambda+3)}\right]\\
&&\hspace{-3.7cm}\langle{{\mathfrak{a}}^{\dag}}^{2}\rangle=\overline{\langle{\mathfrak{a}}^{2}\rangle}\nonumber\\
&&\hspace{-3.7cm}\langle{\mathfrak{a}}^{\dag}\mathfrak{a}\rangle=\sum^{\infty}_{n}\left[(2n)|c_{+,2n}(t)|^{2}+(2n+2\lambda+1)|c_{-,2n+1}(t)|^{2}\right]\\
&&\hspace{-3.7cm}\langle\mathfrak{a}{\mathfrak{a}}^{\dag}\rangle=\sum^{\infty}_{n}\left[(2n+2\lambda+1)|c_{+,2n}(t)|^{2}+(2n+2)|c_{-,2n+1}(t)|^{2}\right]\\
&&\hspace{-3.7cm}\langle1+2\lambda
\hat{R}\rangle=\sum^{\infty}_{n}\left[(1+2\lambda)|c_{+,2n}(t)|^{2}+(1-2\lambda)|c_{-,2n+1}(t)|^{2}\right]\end{eqnarray}
one can derive the variance and covariance of the operators $\hat{x}$ and $\hat{p}$. From Eqs. (29) and (31), we conclude that $S^{\lambda}_{x(p)}$ is strongly dependent on the complex variable $w(= |w| e^{i\phi})$, the deformation parameter $\lambda$, the detuning $\Delta$ and the coupling constant $g$. These dependencies can be discussed as follows:\\
$\bullet$ Figures 8 (a) and (b) visualize variation of the squeezing factors $S_{p}$ and $S_{x}$ in terms of $gt$ for different values of the deformed parameter $\lambda=-0.25, 0$ and $5$ when we choose the phase $\phi=$ 0 and $\frac{\pi}{2}$, respectively. These figures show that the squeezing effect in the field operator $p$ may be considerable for $\phi=0$ and small values of $gt$ while $\lambda>0$. Sometimes, the squeezing factor $S_{p}$ tends to zero, as seen in figure 8(a), which indicates that the states $\Psi(t)$ become minimum uncertainty ones.\\\\
$\bullet$ For the case $\phi=0$, our calculations show that the squeezing factors $S_{p}$ are really dependent of $\lambda$. Figure 8(a) shows that, with increasing of $\lambda$, the degree of squeezing or depth of non-classicality increases at first and then decreases with different $gt$.\\\\
$\bullet$ Squeezing in the $p$ quadrature is disappeared when $\phi$ reaches $\frac{\pi}{2}$, where squeezing in the $x$ quadrature is arised (see figure 8(b)).
\begin{figure}
\begin{center}
\epsfig{figure=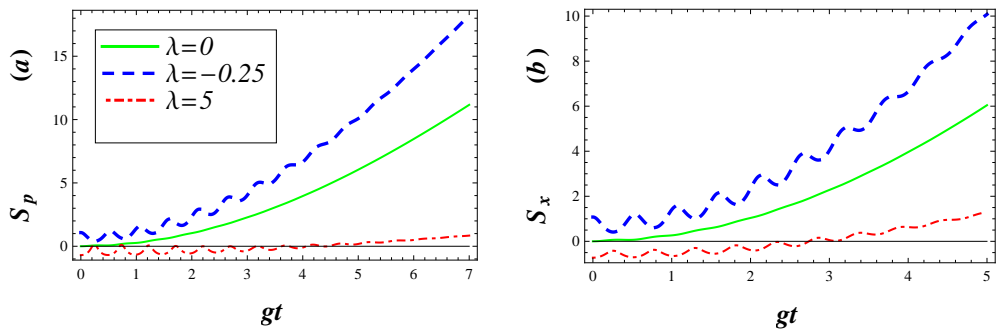,width=15cm}
\end{center}
\caption{\footnotesize Squeezing in the $p$ and $x$ quadratures against $gt$ for different values of $\lambda$ with $g=0.01, |w|^2=9$ and $\Delta=0.1$ as well as for fixed values of $\phi=0$ and $\frac{\pi}{2}$ correspond with (a) and (b) , respectively. The solid curve is plotted for $\lambda= 0$.}
\end{figure}

\section{Conclusions}
A model of deformed Jaynes-Cummings Hamiltonian, expressed in term of fermionic and $\lambda$- deformed bosonic operators, was introduced. It's diagonal form and its eigen-states and eigenvalues were obtained explicitly, analogously to the well known case. Mathematical and physical implications and applications of our results have been also discussed in detail. The deformed JCM introduced, here, could be used to further investigate the interaction between an atomic system and a single mode of an electromagnetic field, including damping or amplifying processes, which are of fundamental importance for example in quantum optics. It was found that the generalized Jaynes- Cummings model is governed by a WHA which reduces to the well-known Heisenberg algebra occurring in the standard JCM. It has been shown that the atomic inversion exhibits Rabi oscillations include quasi-periodic behavior. Searching by the  statistical properties of the deformed JCM reveals that if only the initial field state has a mean photon number exceeding 9 (near a resonant case) significant squeezing can be achieved. It's strength can be arbitrarily large for increasing deformation parameter.  The $\lambda-$deformed JCM can be, potentially, applied to generate maximally entangled states. In other word, the small detuning and coupling regimes with large deformation parameter of this system can support high fidelity and maximally entangled quantum state. Finally, a possible generalization to the three-level system can be discussed.

\end{document}